# Research on Artificial Intelligence Ethics Based on the Evolution of Population Knowledge Base

Feng Liu[1,2]*, Yong Shi [1,2,3,4]*

[1]Research Center on Fictitious Economy and Data Science, the Chinese Academy of Sciences, Beijing 100190, China

[2]The Key Laboratory of Big Data Mining and Knowledge Management Chinese Academy of Sciences, Beijing 100190, China

[3]College of Information Science and Technology University of Nebraska at Omaha, Omaha, NE 68182, USA

[4]School of Economics and Management, University of Chinese Academy of Sciences, Beijing 100190, China

e-mail: liufeng@126.com, yshi@ucas.ac.cn

**Abstract:** The unclear developmentdirection of human society is a deep reason for that it is difficult to form a uniform ethical standard for human society and artificial intelligence. Since the 21st century, the latest advances in the Internet, brain science and artificial intelligence have brought new inspiration to the researchon the developmentdirection of human society. Through the study of the Internet brain model, AI's IQ evaluation, and the evolution of the brain, this paper proposes that the evolution of population knowledge base is the key for judging the development direction of human society,thereby discussing the standards and norms for the construction of artificial intelligence ethics.

## 1. Dilemma for Formation of Human and Artificial Intelligence Ethics

Like the cases of intelligence, consciousness, life and universe, it is difficult to give ethics a uniform definition. Generally,ethics refers to the principles and guidelines that should be followed when one deals with the people-to-people and people-to-society relationships, and it is a philosophical reflection on moral phenomena from a conceptual perspective. Ethics not only contains the behavioral norms in dealing with the people-to-people, people-to-society and people-to-nature relationships but also embodies the profound truth of regulating behaviors in accordance with certain principles[1].

Since the ethics is often related to culture, religion, region, values, and world view, there haven't been a unified, standard, and clear ethical systemin thousands of years of human civilization historyexcept for some principles that are basically acknowledged by people. The ethical impact from the artificial intelligence has become more prominent because of the incompleteness and controversy of ethical issues. The following are several examples related to ethics.

*2012* is a disaster movie about global destruction. In the movie, the president of the USA said "A scientist is more important than dozens of officials." when he let the physicist to leave safely but he himself stayed behind.Which group on earth should get the life priority in a major disaster,has become an important ethical issue in the process of flight.

In the famous ethical thought experiment, i.e.*The Choice of a Switchman*, the train travels at a high speed and couldn't be stopped urgently. Right in front of the train is a forked rail. There are five abductees on the left and one abductee on the right of the rail[2]. Should the switchman choose to let the train go to the left or the right？；

In 2017, Boston Dynamics arranged a scientist to give a test attack to the robot who was carrying boxes. As a result, the robot became unstable and fell down after attacked. The viewers who watched the video through the Internet protested that they violated the rights of robot. In this case,a debate about whether a robot has human rights or not was caused in society.

Behind these problems are profound ethical issues. The final choicewill have an important influence and significance on the ethical construction of artificial intelligence in the future.

## 2. Inspiration of New Technological Advances on the Development Direction of Human Society

Since the 21st century, the Internet, cloud computing, big data, Internet of things, artificial intelligence, brain-like computing, and brain science have emerged. Human technology is experiencing another round of explosive growth. Where, the development of brain science, Internet, and artificial intelligence has provided a new perspective for finding and exploring the development direction of human society.

### 2.1 Evolution of Biological Brain for Hundreds of Millions of Years

For hundreds of millions of years, organisms have followed the principle of "survival of the fittest in natural selection" to form different types of life forms in order to adapt to the changes of the environment. Although themanifestations of organismsvary widely, the core of the organisms,i.e. the brain has shown obvious continuity.From single cell to human, the brain has become more and more complex and the level of intelligence is getting higher and higher.

John. C. Eccles, an Australian scientist, Nobel laureate, mentioned in his book *Evolution of the Brain* that "the brain of organismshas evolved from the brain of fish to the brain of reptiles, then to the brain of mammals and finally to the human brain. If a human brain is dissected, we can see the clear distinction between the structures of fish-likes, reptiles, and mammals in the human brain"[3].

Organisms reflect biological diversity through natural competition and natural selection. Giraffes have longer necks, gazelles run faster, and eagles have sharper eyes. In the structure of the human brain, the biological brain in the process of evolution shows a structure of one layer wrapped by another, just like the accumulated fossil.

## 2.2 Formation and Evolution of the Internet Brain

In 2008, the author of this paper and Professor Peng Geng et al. of the University of Chinese Academy of Sciences published the paper *Trends and Laws of Internet Evolution* by referring to the brain structure of neuroscience based on the emerging Internet-based brain phenomena,and proposed the model of the Internet brain which is used to explain the latest structure of Internet development[4],(see Figure 1).

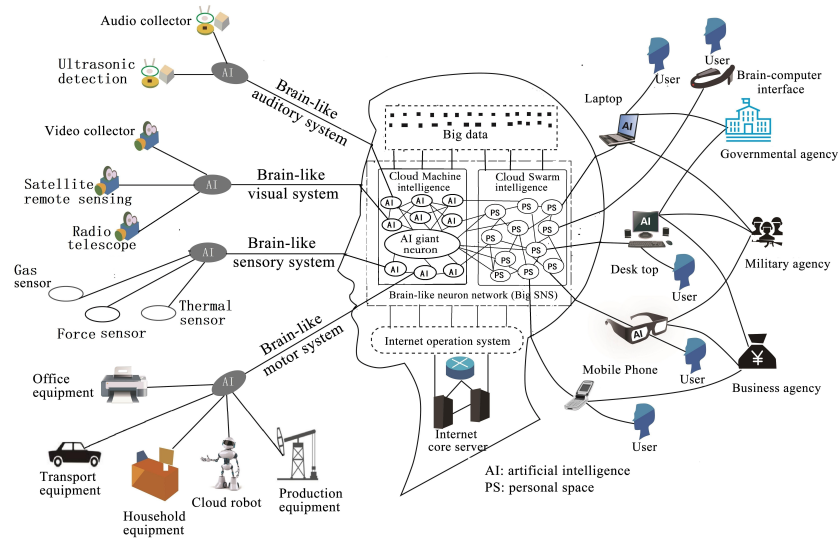

**Figure 1** Architecture of Internet brain

Definition of the Internet brain: The Internet brain is the brain-like giant system architecture formed during the evolution of the Internet to a high-similarity to the human brain. The Internet brain architecture has a brain-like vision system, auditory system, somatosensory system, motor nervous system, memory nervous system, central nervous system and autonomic nervous system. The Internet brain links various social elements (including but not limited to human, AI system, production materials and production tools) and various natural elements (including but not limited to rivers, mountains, animals, plants, and the space) through a neuron-like network.Driven by the collective wisdom and artificial intelligence, the Internet brain achieves the recognition, judgment, decision, feedbackand transformation of the world through the cloud reflex arc.

Regarding the formation of the Internet brain-like giant system, the following can be inferred: When the organisms have evolved to human level, human will be combined with the Internet and achieve co-evolution. And the result of the co-evolution is: The Internet connected with the human is becoming highly similar to the brain step by step in terms of structure, and continuously expands in terms of space along with human expansion(see Figure 2).

## 2.3 Division and Evolution of Intelligence Levels

In 2012, the author of this paper and Professor Shi Yong of the Chinese Academy of Sciences began to consider whether it was possible to evaluate the level of intelligence of the evolving Internet brain. This research topic was later extended to the research on the evaluation of the intelligence level of artificial intelligence systems. The research difficulty is to build a model that can describe uniformly the intelligence characteristics of human life, robots, artificial intelligence systems and the Internet brain.

In 2014, the research was progressed with the reference to the von Neumann architecture, the David Wechsler human intelligence model, and the DIKW model system in the field of knowledge management. A standard intelligence model-Agent was built. It was proposed that any agent, including AI program, robot, human, and the Internet brain model can be described as an integrated system with knowledge input,knowledge mastering, innovation and feedback. The following figure shows the expanded von Neumann architecture added with innovation and cloud storage devices. [5].(see Figure 2).

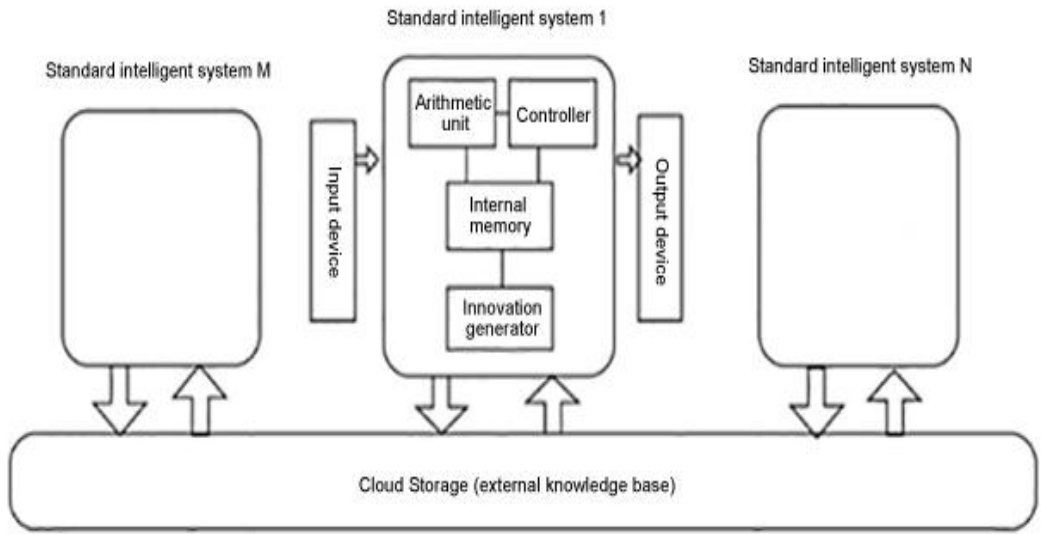

**Figure 2** Expanded von Neumann architecture

The extended von Neumann architecture gives us important inspiration for division of intelligence levels. The criteria for judgment are as follows:

- Can it interact with the tester (human),namely, is there an input/output system?

- Is there a knowledge base in the system that can store information and knowledge?

- Can the knowledge base of this system be continuously updated and increased?

- Can the knowledge base of this system share knowledge with other artificial intelligence systems?

- Can this system actively generate new knowledge and share it with other artificial intelligence systems, in addition to learning and updating its own knowledge base?

In accordance with the above principles, we can have 7 intelligence levels of intelligent systems [5] （Table1）

**Table 1**　Intelligence grades of intelligent systems

| Intelligence Level | Description | Examples |
|---|---|---|
| Level 0 | Intelligent systems in which, for example, the information input can be realized, but the information output cannot. Such intelligent | None |
| | systems exist theoretically, but not in reality. | |
| Level 1 | Intelligent systems that cannot interact with human testers | Stones, sticks, iron pieces and water droplets |
| Level 2 | Intelligent systems that can interact with human testers and have controller and memory, but their internal knowledge base cannot grow. | Floor mopping robots, old-fashioned household refrigerators, air conditioners and washing machines |
| Level 3 | Intelligent systems that have the features of level-2 intelligent systems, and the programs or data contained in the controller or memory of which can be upgraded or added without networking | Smartphones, home computers and stand-alone office software |
| Level 4 | Intelligent systems that have the features of level-3 intelligent systems, what's the most important is that they can share knowledge and information with other intelligent systems through network | Google brain, Baidu brain, cloud robot and B/S architecture websites |
| Level 5 | Intelligent systems that can conduct innovation, recognize and appreciate the value of innovation and creation to human, and apply the results of innovation to the development of intelligent systems | Human |
| Level 6 | Intelligent systems that can constantly innovate and create intelligent systems that generate new knowledge, and their input and output capabilities as well as the abilities to master and use the knowledge will also approach infinity as the time moves forward and approaches to the point of infinity | The "God" of Eastern culture or the concept of "God" in Western culture |

If based on the standard intelligence model, the following mathematical formula can be used to describe the state of omniscience and omnipotence(I-receipt of knowledge information, O-output of knowledge information, S- acquisition or storage of knowledge information, C-innovation of knowledge information)(formula 1)

$$\text{Agent} \rightarrow Q_{\text{Agent}}, Q_{\text{Agent}} = f(\text{Agent})$$

$$Q\text{Agent} = f(\text{Agent}) = f(I,O,S,C) = a*f(I) + b*f(O) + c*f(S) + d*f(C)$$

$$a + b + c + d = 100\%$$

$$f(I) -> \infty, f(O) -> \infty, f(S) -> \infty, f(C) -> \infty$$

**formula 1** the Agent state of omniscience and omnipotence

## 3. Development Direction of Human Society – Judged with the Population knowledge Base

### Inference 1 Population Knowledge Base Evolving toward Omniscience and Omnipotence

The evolution of the brain, the evolution of the Internet, and the division of intellectual levels of intelligence systems all show obvious Increasing attribute. Professor Nelson, an artificial intelligence pioneer, puts forward such a definition of artificial intelligence that: "Artificial intelligence is a discipline of knowledge, and the science about how to express knowledge, acquire knowledge and use knowledge"[6].

A common feature of these three areas is that they are all constantly promoting the loaded knowledge base and ability to use the knowledge.

No matter the evolution of the brain, the evolution of the Internet or the division of intellectual levels of intelligent systems, from the biological development history, it can be seen that the enhancement of knowledge and wisdom is the core of biological evolution. We can judge the direction of biological evolution and the level of biology from the perspective of the capacity of the population knowledge base and the ability to use the population knowledge base(PKB).We can divide PKB into all the biological gene information (GENE) owned by this population, the discovered natural phenomena and the law of operation (SCI), and the grasped technical ability to transform nature (TECH).

When the time approaches infinity, the biological population reaches the "Point of Ω" through the evolution of the population knowledge base toward omniscience and omnipotence.The mathematical formula describing the increment speed of the population knowledge base.(Figure 3)

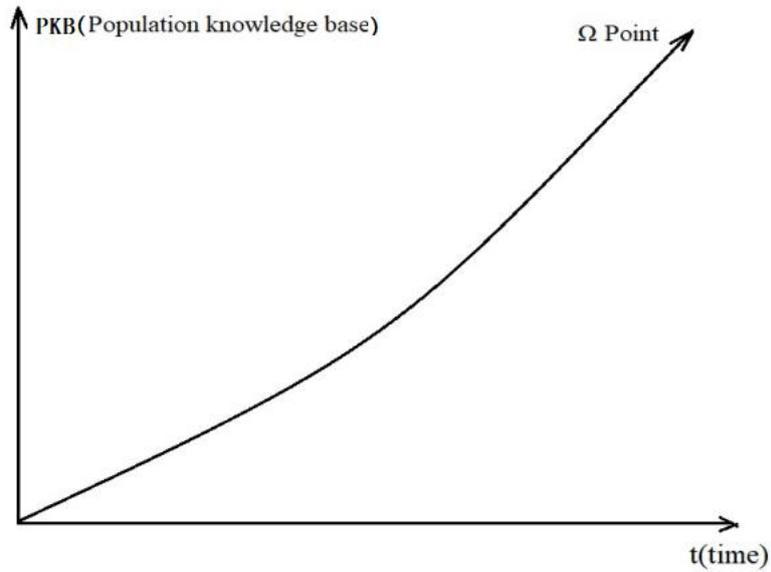

Figure 3 the increment speed of PKB

The mathematical formula describing the increment speed of the population knowledge base.(formula 2)

formula 2    PKB(t)=Gene(t)+Sci(t)+Tech(t)

The omega (Ω) point was proposed by Pierre Teilhard de Chardin (De Rijin), the famous French evolutionist philosopher, in the first half of the 20th century in the book *The Phenomenon of Man*. He used the last Greek letter (Ω) to express this ultimate status of human evolution-omniscience and omnipotence, known as Omega point[7].

**Inference 2 Species Competing with Each Other through the Development of Population Knowledge Base**

The expansion speed of the population knowledge base and the ability to use the population knowledge base is the focus of biological evolution. The knowledge base of other organisms has stagnated and come to a dead end, without further change for thousands of years, so its status in the life circle of the earth is becoming lower and lower.(Figure 4)

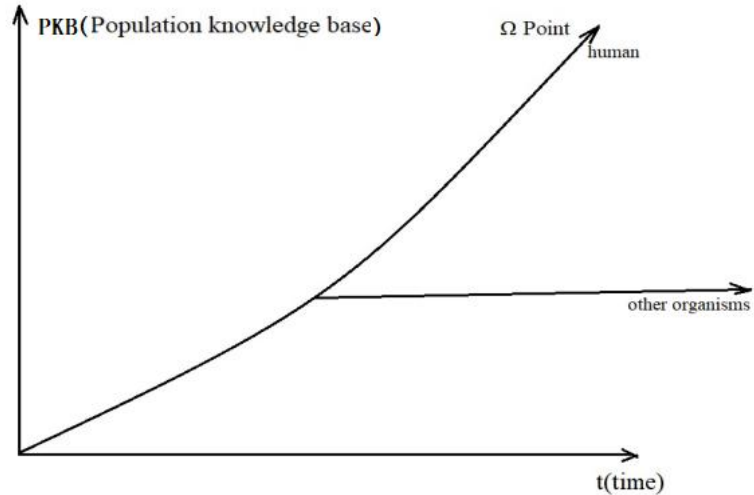

**Figure 4** increment speed of PKB of human and other organisms

In the past 100 thousand years, human has continued to expand and accelerate its knowledge and wisdom, and made further great leaps because of the invention of the Internet and artificial intelligence, thus gaining the dominance in the natural competition of the earth.

$$\text{Formula3} \quad PKB_{human}(t) = gene_{human}(t) + Sci_{human}(t) + tech_{human}(t)$$

$$PKB_{other\ organisms}(t) = gene_{other\ organisms}(t) + Sci_{other\ organisms}(t) + tech_{other\ organisms}(t)$$

## 4. Discussion on the Construction and Decision of AI Ethics from the Perspective of Development Direction of Human Society

4.1 Taking the Evolution from the Population Knowledge Base as a Standard for Constructing AI Ethics

From the discussion in the third section, the promotion of knowledge and wisdom is the core of evolution for human society and even for organisms. We can judge the difference between the development direction of human society and other organisms from the perspective of the capacity of the knowledge base and the ability to use the population knowledge base. When the time approaches infinity, the biological population reaches the "point of God" through the evolution of the population knowledge base toward omniscience and omnipotence. This shows the development of human society has a direction and goal. In the process of life evolution, the behaviors and relationships that promote and protect the development of the population knowledge base are positive ethical rules, while those that hinder and harm the development of the population knowledge base are negative ethical rules.(Figure 5)

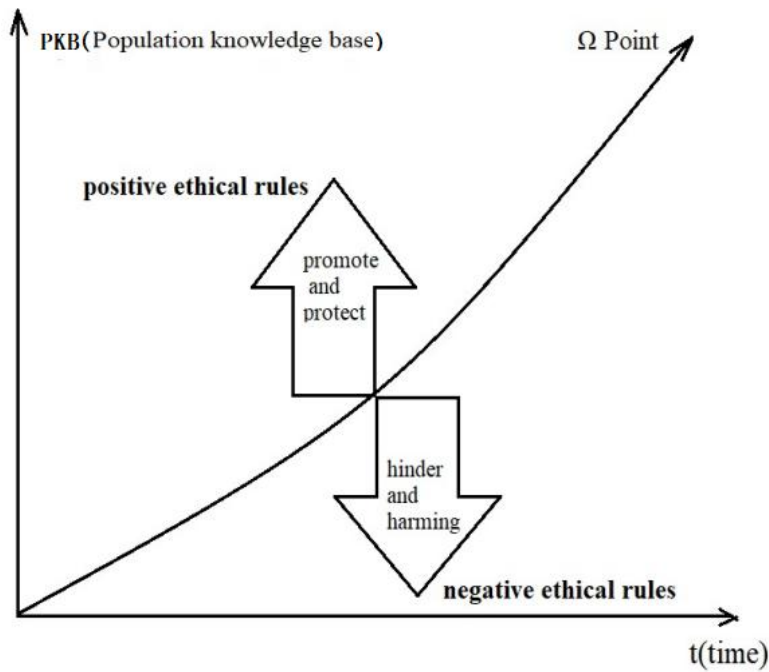

**Figure 5.Relationship betweenPKB and Ethical Rules**

## 4.2 Discussion on the Decision of Viewing AI Ethics from the Perspective of Evolution

1) In the movie of *2012*, the president chose to let the scientist fly to a safe place and expressed that "a scientist is more important than dozens of officials", which was also based on the fact that a scientist is more important than a president to the future of human in terms of continuation and innovation of knowledge.

2) In the *Choice of a Switchman*, can we judge which side contributes more to the future knowledge and wisdom of human? In the absence of a third choice and the ability to judge which side contributes more to the future knowledge and wisdom of human, choosing to allow more people (five) to survive should be a reluctant action.

3) Regarding the ethical issue whether the Boston powered robot has human rights,from the construction of a standard intelligent model and the life evolution mentioned above, AI cannot be seen as a living entity with the same rights as human. It shares part of human knowledge and wisdom functions, but cannot replace human in the most important creativity and evaluation of creative value. More importantly, AI neither can determine its evolutionary direction and evolutionary goals,nor has the natural power for correct evolution.Its evolutionary power comes from human, so it is still one of human's tools(Figure 6).

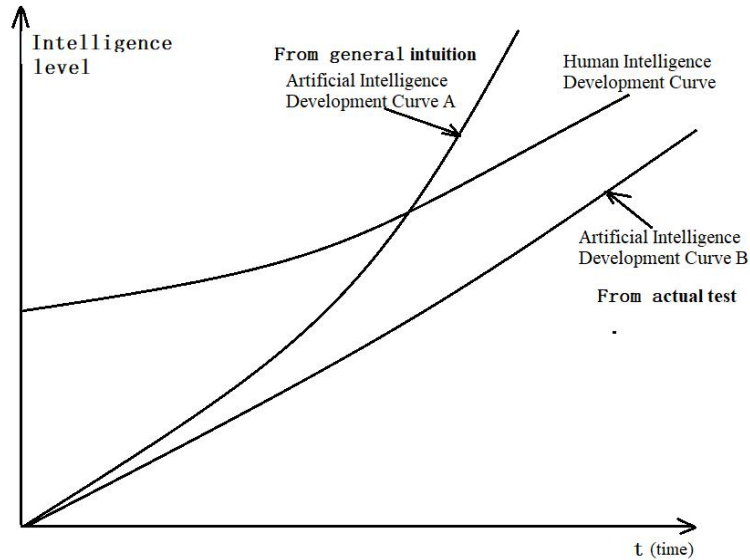

Figure 6. Developmental curves of artificial and human intelligence

4) No matter the great earthquake in *2012*, the escape problem in *Titanic* or the puzzle of the switchman, all belong to the ethical issue concerning escape in extreme situations. However, in the vast majority of cases, individual interests should be protected in non-emergency situations and they should not be occupied by group interests.Allowing individuals to exert their initiatives in exploration should be a very important way for the promotion of the population knowledge base.

5. Summary

It is a new research topic to take the development of population knowledge base as a method for judging the development direction of human society and thereby establish an ethical standard for artificial intelligence,with respect of how the population knowledge base represented by human was formed and developed. If the ethic construction is measured according towhether the development of population knowledge base is promoted or hindered, then how to avoid the mistakes that social Darwinism once committed,and prevent the ethical standard based on population knowledge base from becoming a factor hindering the development of human society will be the worth-studying issues in the future.